\documentclass[pre,aps,onecolumn, superscriptaddress,showkeys]{revtex4}
\usepackage{amsmath}
\usepackage{epsfig}
\usepackage{amssymb}
\usepackage{color}
\usepackage{caption}
\begin{document}

\title{Different measures for characterizing the motion of molecules
  along a temperature gradient}

\author{Oded Farago}
\affiliation{Department of Chemistry, University of Cambridge,
  Lensfield Road, Cambridge CB2 1EW, United Kingdom}
\affiliation{Department of Biomedical Engineering, Ben-Gurion
  University of the Negev, Be'er Sheva 85105, Israel}

\begin{abstract}

We study the motion of a Brownian particle in a medium with
inhomogeneous temperature. In the overdamped regime of low Reynolds
numbers, the probability distribution function (PDF) of the particle
is obtained from the van Kampen diffusion equation
[J. Phys. Chem. Solids {\bf 49}, 673 (1988)]. The thermophoretic
behavior is commonly described by the Soret coefficient - a parameter
which can be calculated from the steady-state PDF. Motivated by recent
advances in experimental methods for observing and analyzing single
nano-particle trajectories, we here consider the time-dependent van
Kampen equation from which the temporal evolution of the PDF of
individual particles can be derived. We analytically calculate the PDF
describing dynamics driven by a generalized thermophoretic force.
Single particles statistics is characterized by measures like the mean
displacement (drift) and the probability difference between moving
along and against the temperature gradient (bias). We demonstrate that
these quantities do not necessarily have the same sign as the Soret
coefficient, which causes ambiguity in the distinction between
thermophilic and thermophobic response (i.e., migration in and against
the direction of the temperature gradient). The different factors
determining the thermophoretic response and their influence on each
measure are discussed.

\end{abstract}
\maketitle

\section{Introduction}

The motion of molecules induced by a temperature gradient is commonly
referred to as thermophoresis, thermodiffusion or the Soret
effect. Since its discovery in liquid mixtures more than a century and
half ago~\cite{ludwig,soret}, the phenomenon of thermophoresis has
been experimentally observed in aqueous solutions containing colloidal
particles, micelles, polymers, proteins, and DNA molecules (see
extensive review in~\cite{piazza08}). Several studies have shown
thermophoresis to be a promising tool for manipulating and
concentrating biomolecules in solutions~\cite{jeon97,duhr06,jiang09},
which has even led to the speculations that it may play a role in the
accumulation of nucleotides required for molecular evolution of early
life~\cite{baaske07}.

In this work we theoretically study the thermal diffusion of colloidal
particles which, in general, is a much stronger effect than
thermophoresis in simple molecular mixtures. The relevant length and
time scales of the colloidal particles are orders of magnitude larger
than those of the embedding solvent, and hence the solvent may be
treated as an effective medium. The thermal motion of the colloidal
particle is driven by stresses induced on its surface by the
surrounding fluid~\cite{ruckenstein81,anderson89,brenner05}.
These forces are balanced by viscous drag forces when the particle
attains a steady state velocity~\cite{burelbach18}. Thermophoresis can
therefore be treated as a mass transport process which, for dilute
suspensions (low concentration, $c$) can be phenomenologically
described by the continuity equation $\partial_t
c=-\overrightarrow{\nabla}\cdot\overrightarrow{J}$, with the particle
flux $\overrightarrow{J}$ given by~\cite{grootmazur}
\begin{eqnarray}
  \overrightarrow{J}=-D\overrightarrow{\nabla}c
  -cD_T\overrightarrow{\nabla}T.
\label{eq:flux}
\end{eqnarray}  
The first term on the r.h.s.~of Eq.~(\ref{eq:flux}) describes regular
diffusion due to concentration gradients, where $D$ is the Fickian
diffusion coefficient. The second term describes an additional
contribution to the flux resulting from the temperature gradient,
$\overrightarrow{\nabla}T$, with $D_T$ termed the thermal diffusion
coefficient. When a closed system reaches a steady state, the flux
vanishes and a concentration gradient is established that satisfies
\begin{eqnarray}
  \overrightarrow{\nabla}c=-cS_T\overrightarrow{\nabla}T,
  \label{eq:soret}
  \end{eqnarray}
where $S_T=D_T/D$ is called the Soret coefficient. For $S_T>0$, the
colloids tend to accumulate on the colder side of the system,
displaying ``thermophobic'' behavior. Conversely, for $S_T<0$, the
migration is toward the hotter side, which is termed ``thermophilic''
motion.

The sign and magnitude of $S_T$ are hard to predict since they depend
on multitude of interactions and influences. Importantly, $S_T$ may
exhibit a pronounced temperature dependence and, quite interestingly,
it tends to change its sign close to room temperature in many
colloidal systems~\cite{iacopini06}. Experimental measurements of
$S_T$ are typically based on the application of a thermal gradient in
a diffusion cell and the use indirect optical methods to quantify the
concentration gradients induced by thermal
diffusion~\cite{piazza08}. Recently, it became possible to measure
thermophoretic forces on a single colloidal particle confined in
sub-micrometer regions with a nearly uniform temperature gradient (and
an overall small temperature difference)~\cite{helden15}. Moreover, we
can now study not only the steady-state probability distribution of
the particle, but to also follow its trajectory to
relaxation~\cite{burelbach17}. These advances in experimental methods
call for a better understanding of the problem of a single particle
diffusion in a temperature gradient.

\section{The van Kampen equation}

Consider a single Brownian particle moving in a one-dimensional medium
with a temperature gradient along the $x$ direction. In order to
derive an equation for the evolution of the probability distribution
function (PDF) of the particle, $P(x,t)$, one has to consider the
Langevin equation of the dynamics or the corresponding Fokker-Planck
equation. These equations capture both the inertial short- and
dissipative long-time regimes of the dynamics. In practice, however,
only the latter is of interest for colloidal systems at low Reynolds
numbers. In this so called ``overdamped limit'', the dynamics is
depicted by a Smoluchowski-like diffusion equation that can be derived
by an adiabatic elimination process of the fast relaxing momentum
degree of freedom. The derivation was carried out by van Kampen for
different models of diffusion in inhomogeneous
media~\cite{vankampen88}. One of the cases considered by van Kampen is
of Brownian particle in a system with spatially-varying
temperature. The equation corresponding to this model is:
\begin{eqnarray}
  \partial_t P(x,t)=-\partial_xJ(x,t)=\partial_x\left\{\mu(x)\partial_x
  \left[k_BT(x)P(x,t)\right]-\mu(x)f(x)P(x,t)\right\},
  \label{eq:vankampen}
\end{eqnarray}
where $f$ is the mechanical force acting on the particle, while $T(x)$
and $\mu(x)$ denote, respectively, the coordinate-dependent
temperature and mobility. The latters are related to the
coordinate-dependent diffusion coefficient, $D(x)$, via Einstein's
relation $D(x)=k_BT(x)\mu(x)$, with $k_B$ denoting the Boltzmann
constant~\cite{vankampen88}. As noted by van Kampen, this is a
diffusion equation which does not follow neither It\^{o}~\cite{ito}
nor Stratonovich~\cite{stratonovich} prescriptions for overdamped
Brownian dynamics in inhomogeneous media.

It is important to note that while the non-isothermal dynamics
considered here is clearly out of thermal equilibrium, the overdamped
limit depicted by van Kampen equation (\ref{eq:vankampen}) is based on
the approximation that the momentum of the particle, $p$, is always at
equilibrium with the local temperature $T(x)$, i.e., follows the
Maxwell-Boltzmann distribution $\rho(p|x)\sim
T(x)^{-1/2}\exp[-p^2/2mk_BT(x)]$ (where $m$ denotes the mass of the
particle). The local thermodynamics equilibrium
(LTE)~\cite{grootmazur} approximation is justified when
$l_b|\overrightarrow{\nabla}T|/T\ll 1$, where $l_b$ is the ballistic
distance characterizing the crossover between the inertial and
diffusive regimes. Mathematically, the overdamped limit corresponds to
$l_b\rightarrow 0$.

The force $f$ in Eq.~(\ref{eq:vankampen}) includes both contributions
from the thermophoretic force, as well externally applied forces like
gravity which can be minimized by density-matching the colloid with
the solvent. We will henceforth ignore all forces except for the
thermophoretic one. Moreover, single particle experiments are
conducted in small systems where the applied temperature difference
may be as small as a few degrees Kelvin. Assuming that the temperature
gradient, $T^{\prime}=dT/dx$, and the thermophoretic force are uniform
throughout the small system, one may phenomenologically write that the
thermophoretic force is given by~\cite{widder89}
\begin{equation}
  f=C_T k_BT^{\prime},
  \label{eq:force}
\end{equation}
where $C_T$ is dimensionless parameter. Using this phenomenological
form in Eq.~(\ref{eq:vankampen}) and comparing with
Eqs.~(\ref{eq:flux}) and (\ref{eq:soret}), we arrive at the following
expression for the Soret coefficient~\cite{fayolle08,sancho18}
\begin{eqnarray}
  S_T=\frac{1-C_T}{T}.
  \label{eq:soret2}
\end{eqnarray}
From Eq.~(\ref{eq:soret2}) we conclude that in the absence of a
mechanical thermophoretic force ($C_T=0$), the Soret coefficient does
not vanish ($S_T\neq 0$). The additional contribution to $S_T$ is
known as the ``ideal gas term''. Explicitly, the $1/T$ term in
Eq.~(\ref{eq:soret2}) is expected because at steady-state
$\partial_tP(x,t)=0$, and from van Kampen equation
(\ref{eq:vankampen}) one can easily deduce that the stationary
solution is
  \begin{equation}
    P_s(x)\sim \frac{1}{T(x)}\exp\left[\int^x\frac{f(y)}{k_BT(y)}\,dy\right],
    \label{eq:sssolution1}
    \end{equation}
which, in the absence of a mechanical force ($f(x)=0$), reduces to
\begin{equation}
  P_s(x)\sim 1/T(x).
  \label{sssolution2}
\end{equation}
In ref.~\cite{vankampen88}, van Kampen notes that he has {\em no
  simple explanation for the prefactor $1/T(x)$}\/ in
Eq.~(\ref{eq:sssolution1}); however, in the special case $f(x)=0$,
Eq.~(\ref{sssolution2}) was nicely rationalized by Fayolle et
al.~\cite{fayolle08}. They noted that the mechanical thermophoretic
force vanishes in the absence of interaction between the colloidal
particles and the embedding solvent, i.e., in the limit of extremely
small colloidal particles that can be viewed as an ideal gas. In a
closed system at steady state, the pressure of this ideal gas,
$\Pi=c(x)k_BT(x)$, must be uniform (or otherwise, the gradient
pressure force would act on the gas and change its
distribution). Eq.~(\ref{sssolution2}) then means that in the absence
of a mechanical thermophoretic force, the ideal gas thermal collisions
induce a steady-state distribution that is higher on the colder than
on the hotter side. The associated Soret coefficient $S_T=1/T>0$
reflects the thermophobic nature of the thermal collisions (which are
stronger on the hotter side, thus ``pushing'' the particle to the
colder side). The Soret effect is associated with the interaction term
in Eq.~(\ref{eq:soret2}) and, in practice, this term is typically
larger than the ideal gas term $|C|\gg 1$, with the exception of
relatively small colloidal particles (see discussion in
Appendix~\ref{sec:appendix}).

Returning to van Kampen equation, we notice that it also takes into
account the spatial variation in the mobility, which within a small
system can be approximated by
\begin{equation}
  \mu(x)\simeq\mu_0+\mu^{\prime}x,
  \label{eq:mobility}
\end{equation}
where $\mu_0$ is the mobility in the middle of the cell at $x=0$ and
$\mu^{\prime}=d\mu/dx$. We note here that the spatial variations in
$\mu(x)$ can, in general, be further divided into two parts - those
arising from the temperature-dependence of the fluid
viscosity~\cite{gutmann52} and those also encountered at equilibrium
isothermal systems, for instance due to hydrodynamic interactions
between the colloidal particle and the walls of the
container~\cite{happelbrenner}. As we will see below, the particle's
drift (to be mathematically defined later) depends on both
$T^{\prime}$ and $\mu^{\prime}$, and one must keep in mind that these
two gradients are not entirely independent of each other because of
the temperature-dependence of the mobility. On the other hand,
non-temperature related reasons for spatial variation in the mobility
imply that $\mu^{\prime}$ does not necessarily vanish when
$T^{\prime}=0$.

Using Eqs.~(\ref{eq:force}) and (\ref{eq:mobility}) [together with the
  expansion $T(x)\simeq T_0+T^{\prime}x$] in the van Kampen equation
(\ref{eq:vankampen}), yields the following form
\begin{eqnarray}
  \partial_t
  P(x,t)=D_0\left\{\partial_x\left(1+
  \frac{x\mu^{\prime}}{\mu_0}\right)\partial_x
  \left(1+\frac{xT^{\prime}}{T_0}\right)
  -C_T\frac{xT^{\prime}}{T_0}\,\partial_x\right\}P(x,t),
  \label{eq:vankampen2}
\end{eqnarray}
where $D_0=D(x=0)=k_BT_0\mu_0$. This is the van Kampen equation in the
limit when (i) all forces besides the thermophoretic one are ignored,
and (ii) the system is sufficiently small to justify the linear
approximations of $T(x)$ and $\mu(x)$. (iii) Another assumption
implied in Eq.~(\ref{eq:vankampen2}) is the form (\ref{eq:force}) for
the thermophoretic force.

\section{The probability distribution function}

In a homogeneous ($\mu^{\prime}=0$) isothermal system
($T^{\prime}=0$), Eq.~(\ref{eq:vankampen2}) reduces to a simple
diffusion equation, the solution of which takes the Gaussian form
$P(x,t)=\exp(-x^2/4D_0t)/\sqrt{4\pi D_0t}\equiv G(x,t)$ [assuming
  Dirac delta-function initial condition $P(x,0)=\delta(x)$]. For a
system under a small temperature difference ($xT^{\prime}/T_0\ll 1$)
and limited changes in the mobility ($x\mu^{\prime}/\mu_0\ll 1$), we
may seek a linear approximate solution of the form
\begin{eqnarray}
P(x,t)=G(x,t)\left[1+x
  H\left(\frac{x^2}{D_0t}\right)\right],
\label{eq:pdf1}
\end{eqnarray}
where $H$ is some function that can be determined in the following
simple manner: (i) Write $H$ in Eq.~(\ref{eq:pdf1}) as a series
expansion in the argument $y=(x^2/D_0t)$: $H=\sum_{i=0}^{\infty}
a_ny^n$, then (ii) determine the coefficients of the expansion by
substituting Eq.~(\ref{eq:pdf1}) in van Kampen
equation~(\ref{eq:vankampen2}), and by comparing terms of similar
order in $y$ on both sides of the equation. In this process of
determining $H(y)$, we ignore the terms that are non-linear in $x$. We
find, $a_0=(-3/4+C_T/2)(T^{\prime}/T_0)-1/4(\mu^{\prime}/\mu_0)$,
$a_1=1/8(T^{\prime}/T_0)+1/8(\mu^{\prime}/\mu_0)$, and $a_n=0$ for
$n>1$, and thus write
\begin{eqnarray}
P(x,t)=\frac{\exp(-x^2/4D_0t)}{\sqrt{4\pi D_0t}}\left[1+x
  \left\{\frac{T^{\prime}}{T_0}\left(-\frac{3}{4}+\frac{C_T}{2}
  +\frac{x^2}{8D_0t}\right)
  +\frac{\mu^{\prime}}{\mu_0}\left(-\frac{1}{4}+\frac{x^2}{8D_0t}\right)
  \right\}\right],
\label{eq:pdf2}
\end{eqnarray}
which is the main result of the paper.

\section{The drift and the flux}

The drift of an individual particle is characterized by the mean
displacement, $\langle x\rangle$, and from the PDF (\ref{eq:pdf2}), we
find that
\begin{eqnarray}
  \langle x\rangle=\int_0^{\infty} xP(x,t)dx=\left(C_T\frac{T^{\prime}}{T_0}
  +\frac{\mu^{\prime}}{\mu_0}\right)D_0t. 
  \label{eq:drift}
\end{eqnarray}
We notice that the drift does not necessarily have the same sign as
$C_T$, which means that the average displacement of the particle is
not necessarily in the same direction as the thermophoretic force. The
reason for this remarkable result is an additional contribution to the
drift originating from spatial dependence of the mobility. In general,
the mobility of simple liquids increases with temperature, while gases
exhibit an opposite trend and have mobility that decreases
approximately like the square root of the
temperature~\cite{gillepsie}. As noted earlier [see discussion after
  Eq.~(\ref{eq:mobility})], non-thermal effects may also contribute to
$\mu^{\prime}$. Indeed, it is well known that drift is also observed
in isothermal systems with non-uniform mobilities~\cite{volpe16}. This
equilibrium phenomenon has been termed ``spurious drift'', which is
misleading since it is a real effect~\cite{farago14}. In the
isothermal case ($T^{\prime}=0$), we can use Einstein relation and
Eq.~(\ref{eq:mobility}) to write Eq.~(\ref{eq:drift}) in the more
common form, $\langle x \rangle/t=D^{\prime}$~\cite{farago14b},
relating the drift velocity and the spatial derivative of the
diffusion coefficient. Thus, our result Eq.~(\ref{eq:drift})
generalizes the well-known expression for the drift of Brownian
particles in isothermal inhomogeneous media to non-isothermal systems.

Recall that the derivation of van Kampen equation (\ref{eq:vankampen})
is based on assuming LTE in the overdamped limit. Within this
approximation, the mean kinetic energy of a particle found at some
coordinate $x$ is related to the local temperature via the
equipartition theorem $\langle E_k\rangle_x=\langle
mv^2/2\rangle_x=k_BT(x)/2$, where $\langle\cdots\rangle_x$ denotes
average at a given $x$. Taking the average with respect to $x$ and
using Eq.~(\ref{eq:drift}) gives
\begin{eqnarray}
  \frac{d\left \langle E_k(t)-E_k(t=0)\right \rangle}{dt}
  =\frac{k_BT^{\prime}}{2}\frac{d\langle x\rangle}{dt}=
  \frac{k_BT^{\prime}}{2}
  \left(C_T\frac{T^{\prime}}{T_0}+
    \frac{\mu^{\prime}}{\mu_0}\right)D_0.
  \label{eq:heat}
\end{eqnarray}
For $\mu^{\prime}=0$ (constant mobility), the particle is heated on
average (i.e., gains kinetic energy) when $C_T>0$, i.e., when the
thermophoretic force drives the particle to the high temperature side,
and vice versa. This, however, may not be true when the mobility
varies in space, in which case it is the sign of
$C_T+(T_0\mu^{\prime}/T^{\prime}\mu_0)$ rather then the direction of
the thermophoretic force that determines whether the particle gains or
losses heat.

A common error is to confuse the above-discussed drift with the flux,
defined by $J(x,t)=-D(x)\partial_xP(x,t)$.  A closed system at steady
state has zero flux, $J=0$, but this does not necessarily imply that
the average displacement (i.e., drift) of each individual particle
must also vanish. On time scales smaller than the characteristic
diffusion time across the system, particles located at different parts
of the system (e.g., near the center or close to the boundaries) may
have different non-vanishing displacements. This situation has been
previously dubbed ``drift without flux'' in equilibrium isothermal
systems~\cite{lancon01}. Here, we consider dynamics in an open system
with time-dependent flux.  The tendency of particles to migrate
favorably to one side may be characterized by the flux at the origin
$J_0\equiv J(x=0,t)=\sqrt{D_0/(\pi
  t)}\left[\left(T^{\prime}/8T_0\right)\left(3-2C_T\right)+\left(\mu^{\prime}/8\mu_0\right)\right]$.
The flux at the origin causes a ``bias'', i.e., a difference in the
probability of finding the particle in the ``hotter'' and ``colder''
sides relative to its initial location. Assuming (without loss of
generality) that $T^{\prime}>0$, the bias, $\Delta (t)$, is defined by
\begin{eqnarray}
  \Delta(t)\equiv\int_0^{\infty} P(x,t)dx-\int_{-\infty}^0
  P(x,t)dx=\sqrt{\frac{D_0t}{\pi}}\left[\frac{T^{\prime}}{2T_0}
   \left(2C_T-1\right)+\frac{\mu^{\prime}}{2\mu_0}\right]
  =\frac{\langle
    x\rangle}{\sqrt{4\pi D_0t}}-\sqrt{\frac{D_0t}{4\pi}}T^{\prime}S_T,
  \label{eq:bias}
\end{eqnarray}
with the drift, $\langle x\rangle$, and the Soret coefficient, $S_T$,
given by Eqs.~(\ref{eq:drift}) and (\ref{eq:soret2}),
respectively. Depending on the values of
$T_0\mu^{\prime}/T^{\prime}\mu_0$ and $C_T$, it now becomes clear that
while $\Delta$, $\langle x\rangle$, and $-S_T$ can all be used to
characterize the response of colloidal particles to a temperature
gradient, these quantities describe different features of the Soret
effect, and may occasionally have different signs.

\section{Discussion and summary}

Motivated by recent single-molecule experiments for studying the
behavior of macromolecules along a temperature gradient, we considered
here the question of Brownian dynamics of a colloidal particle in a
non-isothermal fluid. In the overdamped limit, the PDF of the particle
is described by time-dependent van Kampen diffusion equation
(\ref{eq:vankampen}). Assuming a small temperature and mobility
differences between the ends of the (small) system
($T^{\prime}x/T_0\ll1$ and $\mu^{\prime}x/\mu_0\ll1$), we considered
the linear (in $x$) version of van Kampen equation
(\ref{eq:vankampen2}) and analytically derived the solution for
delta-function initial condition (\ref{eq:pdf2}). The asymmetric PDF
characterizes the general tendency of the particle to migrate in the
direction of the thermophoretic force caused by the temperature
gradient. However, the thermophoretic force is not the only factor
determining the direction of the motion, and we have identified three
different measures for the thermodiffusive response of the colloidal
particle. The first measure is the Soret coefficient $S_T$
(\ref{eq:soret2}), relating the concentration and temperature
gradients in steady state. The Soret coefficient has been
traditionally used to distinguish between thermophilic ($-S_T>0$) and
thermophobic ($-S_T<0$) behaviors. However, we see from
Eq.~(\ref{eq:soret2}) that $-S_T$ and $C_T$ do not necessarily have
the same sign, indicating that the steady-state concentration gradient
is not solely dictated by the direction of the thermophoretic
force. The origin of the discrepancy are the thermal collisions which
set a concentration gradient opposite to the temperature gradient. In
fact, in some recent experiments on colloidal systems it has been
found that $S_T$ exhibits a strong temperature-dependence and tends to
change its sign in the vicinity of room temperature. Moreover, the
magnitude of the Soret coefficient in many of these experiments is
found to be of the order of
$0.01-1K^{-1}$~\cite{iacopini06,braibanti08}. These findings indicate
that (i) the effect of the thermal collisions may sometimes be as
important the thermophoretic force that accounts for the
particle-solvent interactions, and that (ii) the thermophoretic force
(coefficient $C_T$) is sensitive to temperature variations. Due to the
system-specific nature of the thermophoretic force, there is no clear
explanation for its temperature sensitivity of $C_T$ which is likely
dependent on numerous factors, e.g., the thermal expansivity of the
solvent~\cite{iacopini06}, the surface
functionality~\cite{burelbach17} and size~\cite{duhr06} of the
colloidal particle, and electrostatic effects~\cite{putnam07}. In
order to understand this behavior of $C_T$ one must consider a
microscopic model that takes into account some of these factor (see,
e.g., the theoretical discussion in ref.~\cite{sancho18}). This is
beyond the scope of the phenomenological discussion presented herein;
however, in light of the pronounced temperature-dependence of $S_T$,
it must be reemphasized that our derivation assumes that the
thermophoretic force is phenomenologically given by
Eq.~(\ref{eq:force}), namely assuming non-equilibrium linear-response.
The same linear form has been considered in other works (see,
e.g.,~\cite{fayolle08}), and it is consistent with the linearity of
our solution for the PDF (\ref{eq:pdf2}) with respect to
$T^{\prime}$. More generally, the variations of $C_T$ with $T$ can be
accounted for by a Taylor expansion around $T_0$:
$C_T=C_T(T_0)+(dC_T/dT)\Delta T+\cdots=
C_T(T_0)+(dC_T/dT)T^{\prime}\Delta x+\cdots $, which shows that the
linear approximation is valid if the total temperature difference
across the experimental cell $\Delta T=T^{\prime}\Delta x$ is
sufficiently small, i.e., if the the size of the experimental setup,
$\Delta x$, and the temperature gradient, $T^{\prime}$, satisfy
\begin{equation}
    \Delta T=T^{\prime}\Delta x\ll\left|
    \frac{C_T}{dC_T/dT}\right|\approx\left|
    \frac{S_T}{dS_T/dT}\right|.
    \label{eq:taylor}
\end{equation}
In Appendix~\ref{sec:appendix} we review some experimental
measurements of the Soret coefficient where the total temperature
variation $\Delta T$ does not exceed a few degrees Kelvin and, thus,
reasonably satisfy the above criterion.

The second quantity that can be used to characterize thermophoretic
response is the drift of individual particles $\langle x\rangle$
(\ref{eq:drift}), or better, the drift velocity $v=d\langle
x\rangle/dt$. This measure is interesting for two reasons.
First, we now have the experimental means to measure single particle
trajectories. Second, in the overdamped limit, the drift velocity is
directly related to the rate of heat taken from the solvent by the
particle [see Eq.~(\ref{eq:heat})]. Similarly to $-S_T$, a positive
(negative) value of $v/T^{\prime}$ indicates thermophilic
(thermophobic) response, but these quantities are different as
apparent from the comparison of Eqs.~(\ref{eq:soret2}) and
(\ref{eq:drift}). Importantly, the direction of the drift is set by
both directions of the thermophoretic force and the direction of the
mobility gradient. Obviously, part of the mobility spatial variation
can be attributed to the temperature gradient, but it is important to
recall that coordinate-dependent mobility, $\mu(x)$, is also
encountered in isothermal systems, i.e., in equilibrium
situations. Indeed, our result Eq.~(\ref{eq:drift}) generalizes the
expression for the drift velocity in inhomogeneous isothermal
solutions.

Also suggested by Eq.~(\ref{eq:drift}) is that for $\mu^{\prime}=0$,
the temperature gradient causes a non-vanishing drift vanishing drift
only when $C_T\neq0$, i.e., only in the presence of a thermophoretic
force, but not due to thermal collisions (fluctuations) that are also
influenced by the temperature gradient. This can be understood by
noting that the stochastic noise term in the Langevin equation
depicting the dynamics of the particle has a zero mean, even for
multiplicative (state-dependent) noise (see discussion
in~\cite{farago14b}).

Finally, the third quantity defined here is the bias $\Delta$
(\ref{eq:bias}), measuring the probability difference of moving along
and against the temperature gradient. Similarly to the previously
discussed measures, a positive (negative) value of $\Delta/T^{\prime}$
may indicate thermophilic (thermophobic) response. From
Eq.~(\ref{eq:bias}) we infer that the bias may be expressed as a
linear combination of $\langle x\rangle$ and $-S_T$ and, thus, the
value of this quantity is influenced by all three factors of asymmetry
discussed in the work, namely the thermophoretic force, the
spatial-dependence of the mobility, and thermal collision effect.

{\bf Acknowledgments.} I thank Daan Frenkel for numerous insightful
discussions and comments on the topic. This work was supported by the
Israel Science Foundation (ISF) through Grant No. 991/17.

\appendix

\section{Analysis of experimental data}
\label{sec:appendix}

We begin by noting that a key assumption in our theoretical analysis
is the form of Eq.~(\ref{eq:force}), stating a linear relationship
between the thermophoretic force and the temperature gradient. This
form is consistent with the frequently used linear-response theory for
non-equilibrium systems. As discussed in the main text, the strong
variations of $S_T$ with temperature reported in many experimental
studies~\cite{iacopini06,braibanti08,helden15} restrict the validity
of the linear form Eq.~(\ref{eq:soret}) to small systems where the
total temperature difference, $\Delta T$, applied across the
experimental setup satisfy criterion (\ref{eq:taylor}). Reviewing the
experimental data, it can be concluded that the linear approximation
holds reasonably well in many setups where $\Delta T$ does not exceed
a few degrees Kelvin.  (Some noticeable exceptions: (i)
Ref.~\cite{burelbach17} where the $\Delta T$ was as high as 30 K, but
in that work $S_T$ was found to be temperature-independent. (ii) The
measurements of $S_T$ for large colloidal particles of size $2.5\times
10^{-1}$ $\mu{\rm m}$ reported in~\cite{braibanti08} exhibiting
exceptionally strong variations in $S_T$ over a temperature range
smaller than $5K$ which, in fact, calls for care in the interpretation
of the experimental data.)

The distance, $\Delta x$, across which the temperature difference,
$\Delta T$ (of order of a few degrees Kelvin), is applied, varies from
$h\sim 500$ $\mu{\rm m}$ in older experiments~\cite{braibanti08} to
$h\sim 10$ $\mu{\rm m}$ in more recent
ones~\cite{helden15,burelbach17}. Thus, the experimental range of the
temperature gradient is roughly $3\times10^{-3}-3\times10^{-1}$. As
these experiments are conducted around room temperature $T_0\simeq
300K$, we find that $l_T^{-1}\equiv T^{\prime}/T_0\sim
10^{-5}-10^{-3}$ $\mu{\rm m}^{-1}$.  Furthermore, the range of
experimental values for the Soret coefficient varies from $|S_T|\sim
10^{-2} K^{-1}$ for micellar solution, globular proteins and small
colloidal particles ($a\sim 10^{-2}$ $\mu{\rm m}$)~\cite{piazza08} to
$|S_T|\sim 1 K^{-1}$ for large colloidal particles ($a\sim 2.5\times
10^{-1}$ $\mu{\rm m}$)~\cite{braibanti08}. (A noticeable exception is
ref.~\cite{helden15} where $|S_T|\sim 50K^{-1}$ was measured for large
colloidal particles of diameter $a\sim 2.5$ $\mu{\rm m}$.)  Recalling
that $C_T\simeq (1-T_0S_T )$~(\ref{eq:soret2}), we can deduce from
this relationship that the experimental range of the thermophoretic
force coefficient is $-10^3\lesssim C_T\lesssim 10^3$.

The confinement of the particle in a thin slit between two plates
leads to strong variations in the mobility due to hydrodynamic
interactions between the Brownian particle and the walls of the
cell. The hydrodynamic effect overshadows the additional
(non-equilibrium) contribution to $\mu^{\prime}$ due to the
temperature variation which is typically negligible because of the
smallness of $\Delta T$. From theoretical
considerations~\cite{happelbrenner,benesch03} we can estimate that the
relative variations in the mobility, $\Delta \mu/\mu_0\sim a/h$, where
$a$ is the diameter of the colloidal particle. Therefore, the inverse
length $l_{\mu}^{-1}\equiv\mu^{\prime}/\mu_0\sim
a/h^2$. Experimentally, colloids of diameter $a\sim 2.5\times
10^{-2}-2.5$ $\mu{\rm m}$ have been studies, corresponding to a wide
range of values $l_{\mu}^{-1}\sim 10^{-6}-10^{-2}$ $\mu{\rm m}^{-1}$.

Three quantities that characterize the thermophoretic response of a
system are highlighted in the manuscript: $S_T$, $v$ (the drift
velocity), and $\Delta$ (the probability bias). These can be rescaled
to allow direct comparison with $C_T$. We thus define the following
dimensionless quantities
\begin{enumerate}
  \item The scaled negative Soret coefficient, $-\tilde{S_T}\equiv
    -T_0S_T=C_T-1$.
  \item The scaled drift velocity, $\tilde{v}\equiv
    v(T_0/D_0T^{\prime})=C_T+l_T/l_{\mu}$.
    \item The scaled bias, $\tilde{\Delta}\equiv
      \Delta\sqrt{\pi/D_0t}\,(T_0/T^{\prime})=C_T-1/2+l_T/2l_{\mu}$,
\end{enumerate}
where the length scales $l_T$ and $l_{\mu}$ were defined in the
previous two paragraphs.  All of these quantities have the form
$Q=C_T+A$, implying that they do not change sign at exactly the same
temperature like the thermophoretic force coefficient $C_T$. As
discussed extensively in the manuscript, the additional contribution
to each quantity, $A$, arises from both a thermal collision effect
(which is represented by the negative constants in the definitions of
the scaled quantities) and from spatial variations in the mobility
(the terms proportional to $l_{\mu}^{-1}$). Let us look at a few
experimental examples in order to assess the relative importance of
the additional contribution, $A/C_T$, to the thermophoretic force.
\begin{enumerate}

\item In experiments with charged micelles~\cite{iacopini06}, the
  thermophoretic force coefficient was found to be of order
  $|C_T|\lesssim 10$ within the experimental temperature range
  ($|S_T|\sim 10^{-2}$ $K^{-1}$). The size of these micelles is of
  order of a few tens of nanometers, and the cell size in the
  experiments $h>100$ $\mu{\rm m}$. Thus, $l_T^{-1}\sim 10^{-2}$
  $\mu{\rm m}^{-1}$, while $l_{\mu}^{-1}\sim 10^{-5}$ $\mu{\rm
    m}^{-1}$. We therefore conclude that in these classical
  experiments, the hydrodynamics effect is negligible, while the
  thermal collision effect is small but, nevertheless, important
  because the thermophoretic force is also fairly small.
\item When colloidal particles of diameter $a\sim 5\times 10^{-2}$
  $\mu{\rm m}$ are studied in similar diffusion cells, the
  thermophoretic force coefficient is typically an order of magnitude
  larger, $C_T\sim 10^2$~\cite{iacopini06}. For larger colloidal
  particles of size $a\sim 2\times 10^{-1}$ $\mu{\rm m}$, $C_T\sim
  10^3$~\cite{braibanti08}. Thus, in these experiments, the additional
  contributions to the scaled quantities defined above are vanisingly
  small: $A/C_T\ll 1$.
\item Large colloidal particles of size $a\sim 2\times10^{-1}$
  $\mu{\rm m}$ were also studies in ref.~\cite{duhr06}, but in a much
  narrower diffusion cell of height $h\sim 10$ $\mu{\rm m}$. Here we
  also have $C_T\sim10^3$, but in this case $l_T^{-1}\sim 3\times
  10^{-4}$ $\mu{\rm m}^{-1}$ and $l_{\mu}^{-1}\sim 5\times 10^{-3}$
  $\mu{\rm m}^{-1}$. Thus, the sign of the thermophoretic force
  dominates the direction of movement, but the influence of the
  hydrodynamic effect on the drift and the bias may be felt close to
  the transition temperature from thermophilic to thermophobic
  response.
\item In a recent experiment~\cite{burelbach17}, a
  temperature-independent Soret coefficient $S_T\sim 0.2K^{-1}$ was
  measured for colloidal particles of diameter $a\sim 1$ $\mu{\rm m}$,
  diffusing between plates with spacing $h\sim 10$ $\mu{\rm m}$ and an
  unusually large temperature difference $\Delta T\lesssim 30K$. In
  this setup, $C_T\sim 50$, $l_T^{-1}\sim 10^{-2}$ $\mu{\rm m}^{-1}$,
  and $l_{\mu}^{-1}\sim 10^{-2}$ $\mu{\rm m}^{-1}$. These values
  suggest that the thermophoretic force is the key factor in
  determining the diffusive behavior of the colloidal
  particles. Collision and hydrodynamic effects are equally important
  and their influence is about 1-2 orders of magnitude weaker than
  that of the thermophoretic force.
\end{enumerate}

To conclude, in most of the above experimental examples, the magnitude
of $C_T$ is at least one order of magnitude larger than that of other
contributions (denoted collectively by $A$) over most of the
investigated temperature range. Collision effect has influence on the
Soret coefficient of small particles and micelles of diameter not
larger than $5\times 10^{-2}$ $\mu{\rm m}$, especially close to the
transition temperature from thermophilic to thermophobic behavior
(i.e., when $C_T$ becomes small). The hydrodynamic interactions
between the Brownian particle and the walls of the diffusion cell may
influence the drift behavior of large colloidal particles ($a\gtrsim
1$~$\mu{\rm m}$) in small cells ($h\sim 10$~$\mu{\rm m}$).


\end{document}